\documentclass[letterpaper, 12 pt, conference]{ieeeconf}  % Comment this line out
\IEEEoverridecommandlockouts                              % This command is only
\usepackage{graphicx} % for pdf, bitmapped graphics files
\usepackage{bm}
\usepackage{amsmath}
\usepackage[dvipsnames]{xcolor}
\usepackage{soul}
\usepackage{setspace} % for double spacing
\usepackage{soul}

\title{\LARGE \bf Demand Forecasting for Electric Vehicle Charging Stations using Multivariate Time-Series Analysis

}

\author{Saba Sanami, Hesam Mosalli, Yu Yang,  Hen-Geul Yeh, and Amir G. Aghdam % <-this % stops a space
\thanks{Saba Sanami, Hesam Mosalli, and Amir G. Aghdam are with the
Department of Electrical and Computer Engineering, Concordia Univerity,
Montreal, QC, Canada. Emails: saba.sanami@concordia.ca,
hesam.mosalli@mail.concordia.ca,
amir.aghdam@concordia.ca.
Yu Yang is with the Department of Chemical Engineering and Hen-Geul Yeh is with the Department of  Electrical Engineering, California State University Long Beach. Emails: yu.yang@csulb.edu, henry.yeh@csulb.edu.}% <-this % stops a space
}

\begin{document}

\maketitle
\thispagestyle{empty}
\pagestyle{empty}

%%%%%%%%%%%%%%%%%%%%%%%%%%%%%%%%%%%%%%%%%%%%%%%%%%%%%%%%%%%%%%%%%%%%%%%%%%%%%%%%
\begin{abstract} As the number of electric vehicles (EVs) continues to grow, the demand for charging stations is also increasing, leading to challenges such as long wait times and insufficient infrastructure. High-precision forecasting of EV charging demand is crucial for efficient station management, to address some of these challenges. This paper presents an approach to predict the charging demand at 15-minute intervals for the day ahead using a multivariate long short-term memory (LSTM) network with an attention mechanism. Additionally, the model leverages explainable AI techniques to evaluate the influence of various factors on the predictions, including weather conditions, day of the week, month, and any holiday. SHapley Additive exPlanations (SHAP) are used to quantify the contribution of each feature to the final forecast, providing deeper insights into how these factors affect prediction accuracy. As a result, the framework offers enhanced decision-making for infrastructure planning. The efficacy of the proposed method is demonstrated by simulations using the test data collected from the EV charging stations at California State University, Long Beach.

\end{abstract}

%%%%%%%%%%%%%%%%%%%%%%%%%%%%%%%%%%%%%%%%%%%%%%%%%%%%%%%%%%%%%%%%%%%%%%%%%%%%%%%%
\section{INTRODUCTION}

The widespread adoption of electrical vehicles (EVs) is revolutionizing the worldwide transportation industry. This shift is driven by increasing environmental concerns as more people and governments recognize the urgent need to reduce greenhouse gas emissions and combat climate change. Additionally, many governments worldwide are implementing supportive policies, such as subsidies, tax incentives, and investments in charging infrastructure, making it easier for individuals and businesses to transition to EVs. The increase in EV ownership has created a greater need for a reliable charging station network to ensure drivers can conveniently recharge their cars~\cite{liu2012optimal,ye2022learning}.

By 2030, significant investments in EV charging infrastructure are planned across various continents. In Europe, an estimated \$260\ billion will be required to install new charging stations, upgrade power grids, and expand renewable energy capacity. This expansion necessitates the deployment of approximately 2.9 million public charging stations. In North America, the United States is expected to invest at least \$7.5 billion specifically for EV charging infrastructure, with total potential funding reaching up to \$145 billion. These substantial investments reflect the global push to create a robust and extensive network of charging stations to meet the future demand driven by the increasing penetration of EVs~\cite{mckinsey2023netzero}. 

Given this global importance, effectively forecasting the number of charging requests is critical to maintaining a well-functioning and responsive charging infrastructure. Reliable predictions help prevent the overloading of charging stations, reduce wait times for users, and optimize the placement of new stations in areas where demand is anticipated to increase. Moreover, high-precision forecasting enables dynamic pricing strategies that balance demand more efficiently. It also enhances resource management by streamlining staff assignments and maintenance planning, while guiding long-term infrastructure development by revealing key trends in EV usage and energy consumption~\cite{feng2020review,ai2018household,su2022operating}.

There is a rich body of research on demand prediction for EV charging. Traditional methods primarily focus on statistical and parametric approaches, such as the autoregressive integrated moving average model (ARIMA). These methods aim to forecast usage patterns and demand based on historical data~\cite{amini2016arima,yang2023electrical,kim2021forecasting}. Machine learning techniques, on the other hand, provide enhanced prediction accuracy with the use of random forests~\cite{majidpour2016forecasting, lu2018application} and support vector machines (SVM)~\cite{lu2017load}. These methods have been widely used for single-step prediction and often perform poorly in multi-step prediction. Various deep learning approaches have been investigated, including artificial neural networks (ANN)~\cite{zhu2019electric}, convolutional neural networks (CNN)~\cite{zhang2022short} and long short-term memory (LSTM)~\cite{p2023comparative,ma2022multistep}. In addition, several studies have explored hybrid models, which combine different deep learning techniques to leverage the strengths of each method \cite{mohammad2023energy}.

Deep learning models have proved to work well with univariate data for 1-5 hours of multi-step prediction \cite{wang2023short}. However, their performance tends to decrease as the forecast horizon increases when the model is required to predict all future steps at once. In such cases, optimization becomes more difficult because the model has a limited context from which it can learn. As a result, using multivariate time series data, which incorporates additional features, can improve the performance of multi-step prediction.

Given the complexity of EV charging demand patterns, incorporating multiple parameters can enhance multi-step prediction tasks for longer horizons. For example, temperature plays a role, as extreme weather conditions might increase power consumption in an EV. As with some other essential factors, the charging demand can strongly depend on the time of the year and the day of the week. For instance, there are fewer charging requests on university campuses during summer, weekends, or public holidays. Although some researchers have considered these features in predicting the number of requests  ~\cite{unterluggauer2021short, cao2024feature}, it remains open research to discuss the impact of each feature across multiple future time steps in a multivariate framework.

This paper leverages data collected over two years from charging stations at California State University, Long Beach campus, employing relevant features such as temporal variables—weekdays, holidays, and monthly variations—as well as external factors like temperature, which are known to influence charging demand. These features are incorporated into a multivariate LSTM combined with an attention mechanism model to predict the charging demand over a 24-hour horizon at 15-minute intervals. To enhance the interpretability of the model, we apply explainable AI techniques, specifically SHAP (SHapley Additive exPlanations), to elucidate the contribution of each feature to the model’s predictions. The primary contribution of this paper lies in integrating feature-enhanced multivariate time-series analysis with an explainable LSTM-ATT model, offering both accurate multi-step demand forecasts and valuable insights into the factors driving these predictions. This approach not only advances the prediction capabilities for EV charging stations but also provides a transparent framework for understanding the underlying dynamics of the demand.

The rest of the paper is organized as follows. Section~II defines the problem and provides a brief formulation for that. The proposed learning-based method is discussed in Section~III. Comparative simulation results are presented in Section~IV. Finally, Section~V gives the concluding remarks.

\section{Problem Statement}

EV charging stations generate a stream of time-series data, capturing the charging start and end times as well as energy consumption for each vehicle. These data can be processed to count the demand (i.e., the number of charging requests) within any interval. Several numerical, binary, and categorical variables can influence the charging demand. Numerical variables include, among others, environmental conditions, traffic conditions, energy prices, and battery capacity. Binary and categorical variables capture patterns associated with the day of the week, time of the day,  holidays, seasonal trends and type of station.

The key challenges in developing a multivariate forecasting model include (i) managing the propagation of error across a multi-step prediction horizon, (ii) balancing model granularity with computational efficiency, and (iii) identifying and quantifying the contribution of each feature to the final prediction, thereby enhancing model interpretability. Addressing these challenges is essential for building a robust and accurate forecasting tool that can support the efficient management of EV charging infrastructure.

Given a dataset, let $x^i_t$ denote the value of the $i$-th feature at time $t$, $i \in \{1, 2, \ldots, n\}$, where $n$ is the total number of input features defining the structure of the dataset. The feature set comprises a subset of characteristics including ambient temperature, day of the week, month, and holiday status. Notably, $x^1_t$ represents the number of requests at time $t$, i.e., the number of active EV chargings, which is the target to be predicted. The multivariate observation at time $t$ is formalized as $X_t = [x^1_t, x^2_t, \ldots, x^n_t]$.

Given a sequence of $p$ preceding time steps, denoted by $(X_{t-p+1}, \ldots, X_t)$, it is desired to forecast the future values of the EV charging demand over the prediction horizon, denoted by $m$ time steps, specifically $(x^1_{t+1}, \ldots, x^1_{t+m})$. This is performed by training a robust mapping function that can accurately capture the temporal dependencies and feature interactions in the historical data to generate high-precision predictions of future demand.
\section{Methodology}

We use a multivariate LSTM network enhanced with an attention mechanism. The input to the model consists of multivariate time-series data. The LSTM layer processes these sequences, learning patterns over time that contribute to changes in EV charging demand. Using multivariate data, the model considers the combined effects of various factors on the demand, leading to more accurate predictions.

Next, an attention mechanism is applied to dynamically assign weights to the most relevant time steps, allowing the model to focus on temporal aspects. Additionally, the feature-based approach enables one to use SHAP to quantify the contribution of each feature to the model predictions. The details of each model are explained in the sequel.

\subsection{LSTM}
The LSTM network is a type of recurrent neural network (RNN) specifically designed to capture long-term dependencies in sequential data. For the multivariate time series in this study, where each observation at time $t$ is represented by $X_t$, the LSTM processes the sequence of past observations $(X_{t-p+1}, \ldots, X_t)$ to predict the demand in the future time steps. At each time step, the LSTM maintains a hidden state $h_t$ and a cell state $C_t$, updated based on the previous state and the current input $X_t$. The update process is governed by several units of the forget gate, input gate, and output gate. Mathematically, the forget gate $f_t$ is defined as
\[
f_t = \sigma(W_f X_t + U_f h_{t-1} + b_f).
\]
It determines how much of the previous cell state $C_{t-1}$ is retained, where $W_f$, $U_f$ are weight matrices, and $b_f$ is the bias vector, and $\sigma$ is the sigmoid function. The input gate $i_t$ and the candidate cell state $\tilde{C}_t$ work together to update the cell state as
\[
i_t = \sigma(W_i X_t + U_i h_{t-1} + b_i), 
\]
\[
\tilde{C}_t = \tanh(W_C X_t + U_C h_{t-1} + b_C).
\]
The new cell state is then computed as a combination of the old cell state and the updated information
\[
C_t = f_t \odot C_{t-1} + i_t \odot \tilde{C}_t.
\]
Finally, the output gate $o_t$ determines the updated hidden state as follows~\cite{yu2019review}
\[
o_t = \sigma(W_o X_t + U_o h_{t-1} + b_o), \quad h_t = o_t \odot \tanh(C_t).\]
In this way, the LSTM effectively captures both the temporal dependencies and the multivariate feature interactions in the input data, allowing the model to learn complex patterns in EV charging demand.
\subsection{Attention mechanism}
\begin{figure}
    \centering
    \includegraphics[width=0.8\linewidth]{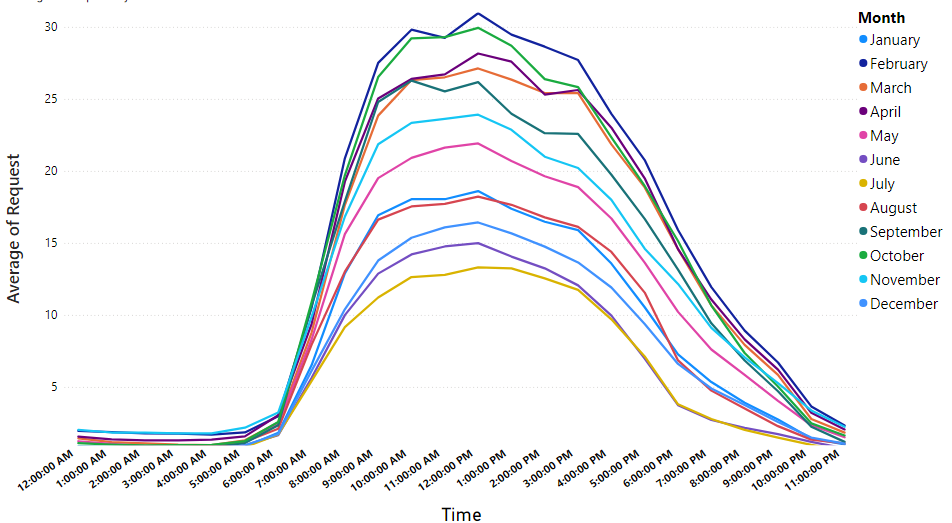}
    \caption{Daily patterns of EV charging requests across different months}
    \label{hour}
\end{figure}

The attention mechanism allows the model to focus on the critical time steps that influence future demand most, improving both the accuracy and interpretability of the predictions. As shown in Fig.~\ref{hour}, the average number of EV charging requests varies significantly at different times of the day. This temporal variation highlights the importance of including a time-step feature in the model, enabling the network to learn patterns based on the time of the day. We incorporate an attention mechanism that assigns weights to different steps to further enhance the model’s ability to focus on the crucial instances and to prioritize the most relevant periods
for prediction.

For a given sequence of past observations $(X_{t-p+1}, \ldots, X_t)$, the attention mechanism computes a set of weights $a_{t}$ for each time step, as
\[
e_t = \tanh(W_a h_t + b_a),
\]
where $W_a$ and $b_a$ are the weight matrix and bias vector for the attention layer, respectively. The attention scores are then normalized using a softmax function to ensure they sum to 1 across all time steps, i.e.,
\[
a_t = \frac{\exp(e_t)}{\sum_{k=t-p+1}^{t} \exp(e_k)}.
\]
The final context vector is computed as the weighted sum of the LSTM hidden states~\cite{vaswani2017attention}
\[
c_t = \sum_{k=t-p+1}^{t} a_k h_k,
\]
which captures the most relevant information from the sequence and is then used to predict future EV charging demand. 

\subsection{SHAP for feature importance}

\begin{figure}
    \centering
    \includegraphics[width=0.8\linewidth]{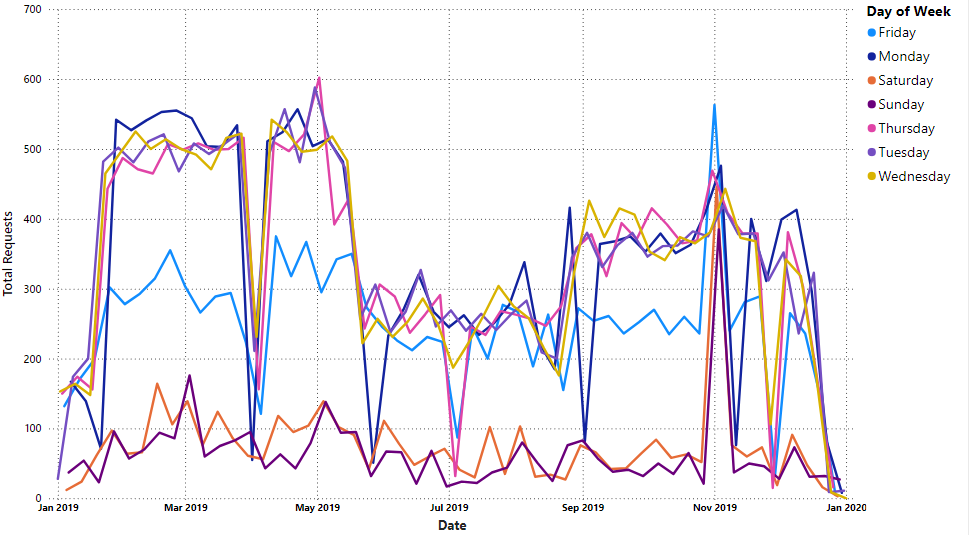}
    \caption{Weekly patterns of EV charging requests over one year}
    \label{pattern}
\end{figure}

As illustrated in Fig.~\ref{pattern}, the charging demand for EVs follows distinct patterns throughout the week. From Monday to Thursday, the pattern remains relatively consistent. However, the demand pattern changes on Fridays, where we observe a slight decline, followed by a more dramatic decrease in charging requests over the weekends. In addition, there is a sudden drop in the number of requests over holidays, reflecting a sharp reduction in charging demand. Another key observation is the seasonal variation, with the summer months experiencing notably fewer requests than other seasons. We apply SHAP to quantify each feature contribution to these patterns and better understand the driving factors behind EV charging demand.

SHAP offers a reliable method for calculating the impact of each feature on the model output by attributing the overall prediction to individual features and considering all possible combinations of feature importance. SHAP values quantify the effect of each feature by determining how the prediction would change if that feature is removed from the model~\cite{salih2024perspective}. Given the multivariate structure of the data, the method allows one to decompose the prediction into contributions from each feature. By doing so, we can identify which features drive the fluctuations in charging demand over time. This feature-based interpretability is critical for making informed decisions about resource allocation at charging stations.

SHAP analysis begins with selecting a set of background data, which serves as a reference for interpreting the model predictions. This baseline represents the typical distribution of feature values on which the model was trained. Background data helps contextualize new predictions by showing how they differ from expected behavior~\cite{holzinger2022explainable}.

\section{Experimental testing of the method}
\subsection{Data collection}
This study collects the actual EV charging data on a university campus in 2019 and 2023. Data for 2020-2022 is excluded due to the campus closure during the COVID-19 pandemic. Each data entry indicates the charging start time, completion time, disconnection time, and the total delivered energy to each car. The charging infrastructure consists of J1772 type-I chargers with a maximum charging power output of 7.2kW. 

\subsection{Data preprocessing}
The original dataset records the times when vehicles are connected to charging stations. To standardize the dataset for analysis, we aggregate the data into uniform 15-minute intervals by counting the number of EVs charging within each interval. This step ensures the data is consistent and can be easily integrated with additional features. Afterwards, we incorporate additional external features into the dataset. We include temperature, an essential factor influencing EV usage, by retrieving 15-minute temperature readings for each timestamp. Additionally, we integrate an academic calendar to account for university-specific events and holidays. Weekday and monthly indicators are also extracted and treated as categorical variables. We then apply one-hot encoding to categorical variables to facilitate using categorical features in time-series forecasting models.

\begin{figure}
    \centering
    \includegraphics[width=0.6\linewidth]{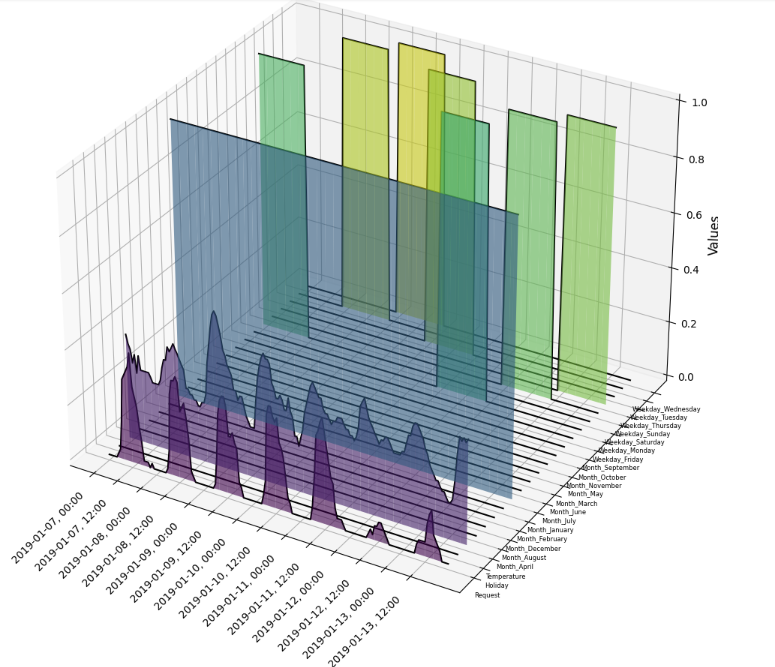}
    \caption{Multivariate data for selected features over one week}
    \label{data}
\end{figure}

The next step involves scaling the data using the MinMax scaler, which normalizes features to a [0, 1] range to ensure consistent contribution to the model's learning process. Fig. \ref{data} represents the normalized data. After scaling, we structure the data for the LSTM-ATT model by creating input-output sequences tailored for time-series forecasting. Each input sequence consists of 96 steps, corresponding to 24 hours of historical data sampled at 15-minute intervals, while the output sequence predicts the next 96 steps, effectively forecasting the subsequent 24 hours of EV charging demand.

The dataset, comprising 70,080 samples collected over two years, is split, with 80\% allocated for training and 20\% for testing. The input data is structured with a shape of (\text{number of samples}, 96, 21), representing 96 past step sequences with 21 features. The output data has a shape of (\text{number of samples}, 96), corresponding to the predicted 96-hour time series.

\subsection{Test results}
The model begins with an input layer that accepts time-series data. The data is then passed to the LSTM layer of 96 units. An attention mechanism is applied on top of the LSTM output to assign importance to different timesteps dynamically. A dense layer computes attention scores for each timestep, highlighting the most relevant periods for predicting future EV charging demand. These attention scores are normalized using a softmax function to create attention probabilities. These probabilities are then multiplied element-wise with the LSTM outputs, weighting the most important timesteps more heavily in the final prediction. The final output layer uses a Dense layer with ReLU activation to produce the forecasts for each 15-minute interval over the 24 hours. Additionally, the model outputs the attention weights, which provide interpretability by indicating which time intervals contributed most to the forecast.  

Tables~\ref{tab:hyperparameters} and ~\ref{tab:performance_metrics} present the model hyperparameter and performance metrics, respectively, indicating that the model predicts unseen data with a reasonably low error rate. Table \ref{tab:model_performance} compares univariate LSTM and multivariate LSTM with and without attention mechanism. As shown, the test loss is higher than the proposed method.

\begin{table}[h]\caption{Model hyperparameters}

\centering
\begin{tabular}{|l|l|}
\hline
\textbf{Hyperparameter}       & \textbf{Value/Description}                                    \\ \hline

\textbf{LSTM Units}           & 96                                                            \\ \hline
\textbf{Attention mechanism}  & Dense layer with 1 unit        \\ \hline
\textbf{Output activation}    & ReLU                                                          \\ \hline
\textbf{Optimizer}            & Adam                                                          \\ \hline
\textbf{Learning rate}        & 0.001                                                         \\ \hline
\textbf{Loss function}        & Mean Squared error                                            \\ \hline

\textbf{Number of epochs}     & 15                                                            \\ \hline
\end{tabular}
\label{tab:hyperparameters}
\end{table}
\begin{table}[h]\caption{Model Performance Metrics}

\centering
\begin{tabular}{|l|l|}
\hline
\textbf{Metric}               & \textbf{Value}           \\ \hline
\textbf{Training loss (MSE)}        & 0.0039                   \\ \hline
\textbf{Test loss (MSE)}            & 0.0085                   \\ \hline
\textbf{Training time}        & 12:24 minutes               \\ \hline
\end{tabular}
\label{tab:performance_metrics}
\end{table}

\begin{table}[h]
\caption{Comparison of model performance with test loss and training time}
\centering
\begin{tabular}{|l|l|p{1.7cm}|}
\hline
\textbf{Model}                & \textbf{Test loss} & \textbf{Training time} \\ \hline
Univariate LSTM               & 0.2124                   & 10:21min                           \\ \hline
Univariate LSTM-ATT           & 0.2014                   & 10.15min                           \\ \hline
Multivariate LSTM             & 0.01178                  & 12:35min                         \\ \hline
Multivariate TCN-ATT          & 0.03456                  & 10:56min                          \\ \hline
\end{tabular}

\label{tab:model_performance}
\end{table}

\begin{figure}
    \centering
    \includegraphics[width=0.8\linewidth]{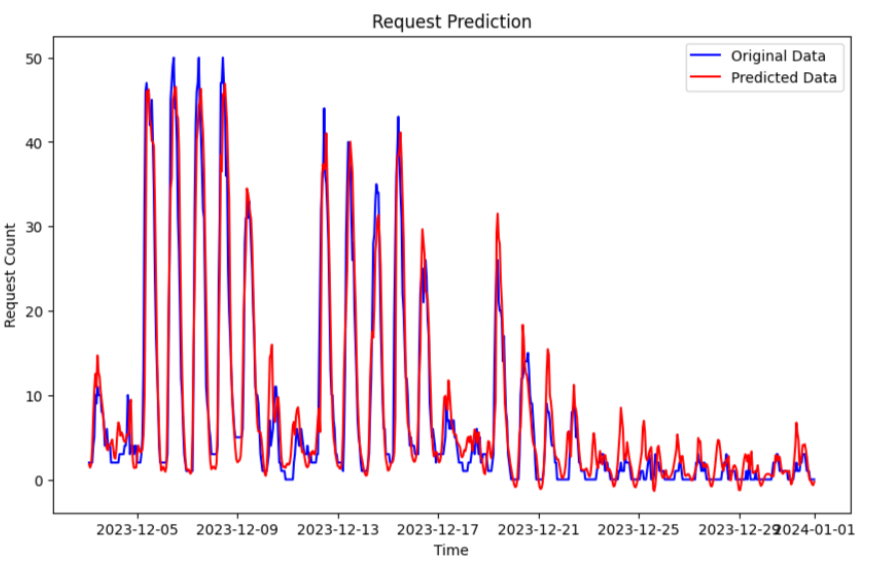}
    \caption{Actual and predicted charging requests}
    \label{real vs predicted}
\end{figure}

\begin{figure}
    \centering
    \includegraphics[width=0.8\linewidth]{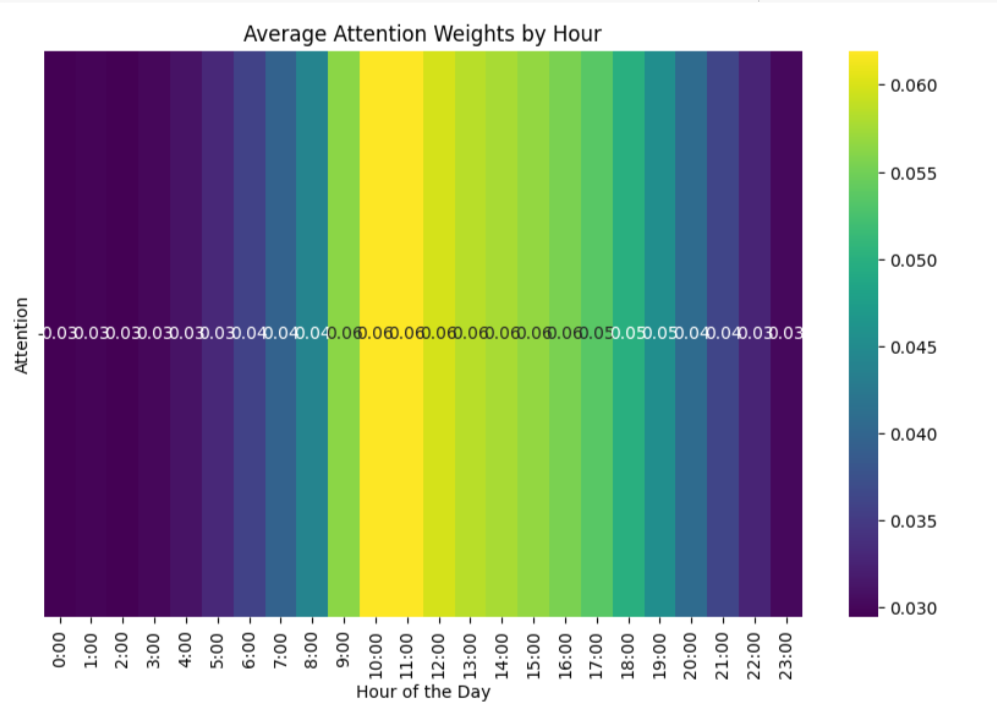}
    \caption{Average attention weights on an hourly basis}
    \label{attention}
\end{figure}

We compare the predicted charging demand with the actual demand over a test period, which includes the end-of-year holidays. Fig.~\ref{real vs predicted} shows the predicted versus actual values, demonstrating that the model accurately captures the overall trend of EV charging demand, with only minor deviations between the predicted and actual values. The model also predicts the demand during the holiday period with high accuracy, which can be attributed to the effective extraction of relevant features.
 The attention mechanism has provided valuable insights into which time steps influence the predictions most. As illustrated in Fig.~\ref{attention}, the attention weights visualization reveals that specific periods, particularly peak hours during the day, are assigned higher weights, indicating their importance in driving the predictions.

\subsection{Explainable AI}

To further understand the contribution of individual features to the model predictions, we use SHAP, whose values can be applied both globally and locally. Global SHAP analysis helps one to identify which features generally have the most influence on the model predictions across the entire dataset. In contrast, local SHAP values allow one to examine the contribution of each feature for a specific instance or test sample.

In the following example, we apply local SHAP to assess the feature contributions for a specific test case.

\begin{figure}
    \centering
    \includegraphics[width=0.8\linewidth]{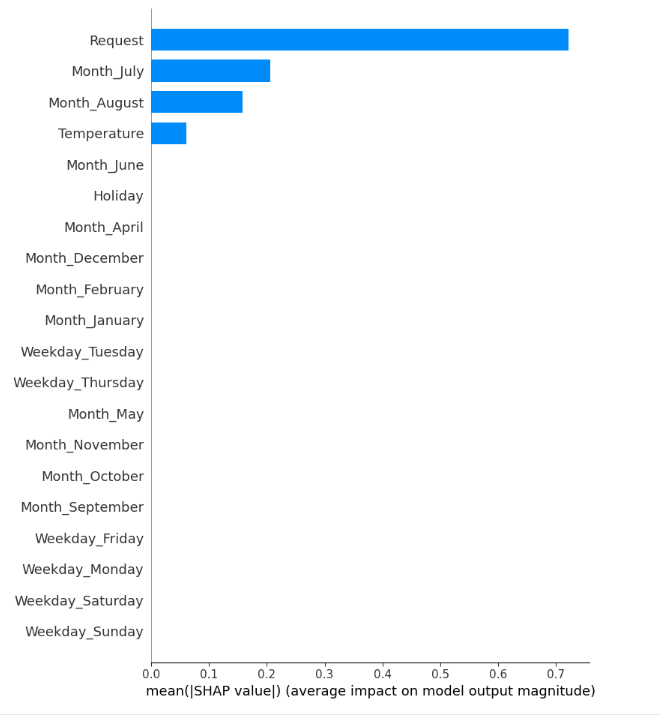}
    \caption{SHAP values in Example~1}
    \label{shap1}
\end{figure}

\textbf{Example 1}. As the first example, we select a Wednesday in July 2023 as the background data and a Wednesday in August 2023 as the test data. The SHAP values shown in Fig.~\ref{shap1} demonstrate that the historical charging request (denoted as Request) is by far the most significant feature influencing the model's predictions. This result is not unexpected, as past demand often serves as a reliable indicator of future demand. The second most influential feature is the month of July, which exerts a substantial impact.

Temperature has a relatively smaller influence on the predictions, which can be attributed to the small temperature difference between the months of July and August. Given that the selected dates are non-holidays, the holidays feature has no significant contribution to the predictions for this particular instance.

\textbf{Example~2}. We apply SHAP to analyze the feature contributions when the background data is taken from a Wednesday in January and the test data from a Wednesday in August. Both dates are non-holidays, allowing one to focus on how other features, such as temperature and the month, influence the model predictions.

\begin{figure}
    \centering
    \includegraphics[width=0.8\linewidth]{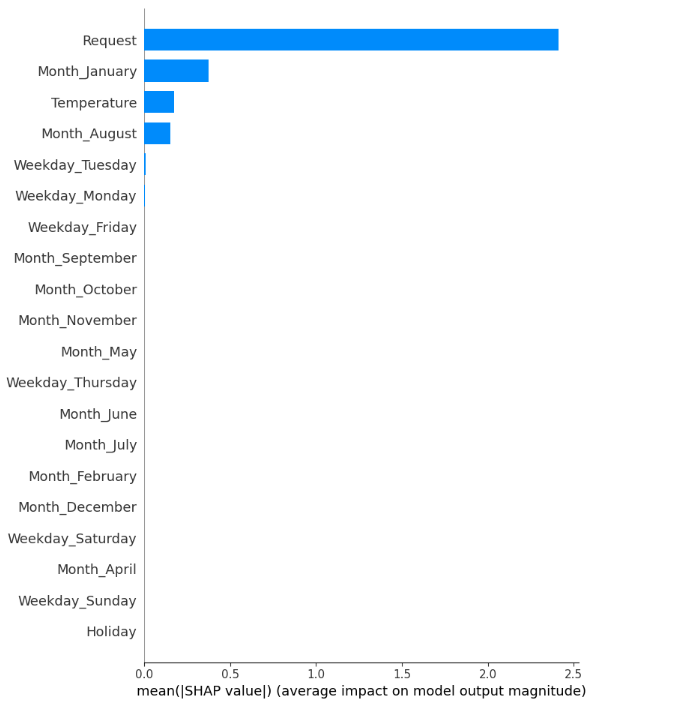}
    \caption{ShAP values in Example~2}
    \label{shap-janaug}
\end{figure}

In Fig.~\ref{shap-janaug}, we observe that the historical charging request continues to be the most significant feature, consistent with the previous example. However, in this case, the temperature plays a more substantial role than the month since the temperature difference between January and August is much more significant, leading to greater variations in charging demand. The increase in temperature during August likely results in higher power consumption due to air conditioning, causing an increase in the number of requests compared to January.

On the other hand, the month itself has a lower impact than the temperature, as the seasonal variation (summer versus winter) has already been captured by the temperature feature. The weekday feature (e.g., Monday, Tuesday, etc. ) also has a minimal influence in this specific case, as both dates fall on the same day of the week.\\

\begin{figure}
    \centering
    \includegraphics[width=0.8\linewidth, trim={0pt 0pt 0pt 10pt},clip]{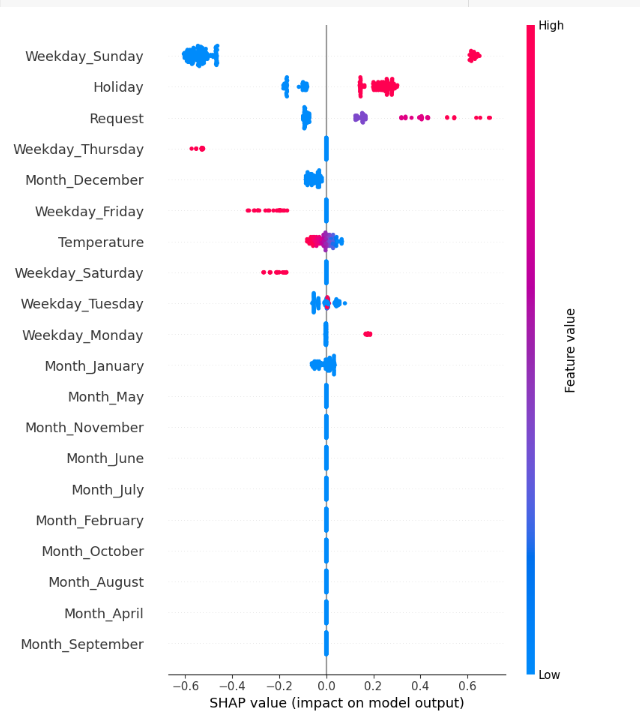}
    \caption{SHAP values in Example~3 using a beeswarm plot}
    \label{beeswarm}
\end{figure}

\textbf{Example~3}. We analyze a series of data from December and January, which includes various weekdays and holidays, to explore how each of these features contributes to the model predictions. We present the results in a SHAP beeswarm chart, as shown in Fig.~\ref{beeswarm}, where each dot represents a SHAP value for an individual prediction. The color gradient represents the feature value (from low to high), and the SHAP value on the x-axis indicates the feature’s contribution to the model output.

In this example, the most important feature is Sunday. A higher value for Sunday (which is binary and takes the value of 1 when the day is Sunday) leads to an increase in the predicted number of requests for the following day, which is Monday. This result is intuitive, as Sundays typically have fewer requests, and the next day (Monday) sees a rise in charging demand.

Moreover, holidays contribute significantly during December. A higher value for this feature results in increased predictions for the following day’s requests. This makes sense, as the holiday period generally sees a surge in activity the day after, especially in terms of travel and vehicle use.

On the other hand, the effect of Thursday, Friday, and Saturday is the reverse of Sunday. Higher values for these days (indicating that the day is either Thursday, Friday, or Saturday) yield fewer requests on the following day. This is consistent with what we observe in real-world patterns, where the number of EV charging requests tends to decrease over weekends and picks up again at the start of the week. SHAP values help explain how these weekdays contribute to the model's lower prediction for the following day.
\section{Conclusions}
We present a feature-enhanced multivariate LSTM model with attention to predict the number of EV charging station requests over the next day at 15-minute intervals. Our results demonstrate that this method achieves higher accuracy compared to models using univariate signals. Additionally, we analyze and explain the contribution of each feature to the final prediction across three different case studies. For future research, we plan to incorporate additional factors into the model, such as battery lifespan, the growing number of EVs, and the collaborative behavior of stations within a given area.

\bibliographystyle{IEEEtran} % This specifies the style in which the references are formatted. Popular styles include `plain`, `unsrt`, `alpha`, and `abbrv`.

\bibliography{reference}

\begin{thebibliography}{10}
\providecommand{\url}[1]{#1}
\csname url@samestyle\endcsname
\providecommand{\newblock}{\relax}
\providecommand{\bibinfo}[2]{#2}
\providecommand{\BIBentrySTDinterwordspacing}{\spaceskip=0pt\relax}
\providecommand{\BIBentryALTinterwordstretchfactor}{4}
\providecommand{\BIBentryALTinterwordspacing}{\spaceskip=\fontdimen2\font plus
\BIBentryALTinterwordstretchfactor\fontdimen3\font minus
  \fontdimen4\font\relax}
\providecommand{\BIBforeignlanguage}[2]{{%
\expandafter\ifx\csname l@#1\endcsname\relax
\typeout{** WARNING: IEEEtran.bst: No hyphenation pattern has been}%
\typeout{** loaded for the language `#1'. Using the pattern for}%
\typeout{** the default language instead.}%
\else
\language=\csname l@#1\endcsname
\fi
#2}}
\providecommand{\BIBdecl}{\relax}
\BIBdecl

\bibitem{liu2012optimal}
Z.~Liu, F.~Wen, and G.~Ledwich, ``Optimal planning of electric-vehicle charging
  stations in distribution systems,'' \emph{IEEE Transactions on Power
  Delivery}, vol.~28, no.~1, pp. 102--110, 2012.

\bibitem{ye2022learning}
Z.~Ye, Y.~Gao, and N.~Yu, ``Learning to operate an electric vehicle charging
  station considering vehicle-grid integration,'' \emph{IEEE Transactions on
  Smart Grid}, vol.~13, no.~4, pp. 3038--3048, 2022.

\bibitem{mckinsey2023netzero}
{McKinsey \& Company}, ``The net-zero materials transition: Implications for
  global supply chains,''
  \url{https://www.mckinsey.com/industries/metals-and-mining/our-insights/the-net-zero-materials-transition-implications-for-global-supply-chains},
  2023, [Online].

\bibitem{feng2020review}
H.~J. Feng, L.~C. Xi, Y.~Z. Jun, Y.~X. Ling, and H.~Jun, ``Review of electric
  vehicle charging demand forecasting based on multi-source data,'' in
  \emph{Proceeding 2020 IEEE Sustainable Power and Energy Conference (iSPEC)},
  2020, pp. 139--146.

\bibitem{ai2018household}
S.~Ai, A.~Chakravorty, and C.~Rong, ``Household {EV} charging demand prediction
  using machine and ensemble learning,'' in \emph{Proceeding 2018 IEEE
  International Conference on Energy Internet (ICEI)}, 2018, pp. 163--168.

\bibitem{su2022operating}
S.~Su, Y.~Li, Q.~Chen, M.~Xia, K.~Yamashita, and J.~Jurasz, ``Operating status
  prediction model at {EV} charging stations with fusing spatiotemporal graph
  convolutional network,'' \emph{IEEE Transactions on Transportation
  Electrification}, vol.~9, no.~1, pp. 114--129, 2022.

\bibitem{amini2016arima}
M.~H. Amini, A.~Kargarian, and O.~Karabasoglu, ``Arima-based decoupled time
  series forecasting of electric vehicle charging demand for stochastic power
  system operation,'' \emph{Electric Power Systems Research}, vol. 140, pp.
  378--390, 2016.

\bibitem{yang2023electrical}
Y.~Yang and H.-G. Yeh, ``\BIBforeignlanguage{English}{Electrical {Vehicle}
  {Charging} {Infrastructure} {Design} and {Operations}},'' no. 23-14, Jul.
  2023.

\bibitem{kim2021forecasting}
Y.~Kim and S.~Kim, ``Forecasting charging demand of electric vehicles using
  time-series models,'' \emph{Energies}, vol.~14, no.~5, p. 1487, 2021.

\bibitem{majidpour2016forecasting}
M.~Majidpour, C.~Qiu, P.~Chu, H.~R. Pota, and R.~Gadh, ``Forecasting the {EV}
  charging load based on customer profile or station measurement?''
  \emph{Applied energy}, vol. 163, pp. 134--141, 2016.

\bibitem{lu2018application}
Y.~Lu, Y.~Li, D.~Xie, E.~Wei, X.~Bao, H.~Chen, and X.~Zhong, ``The application
  of improved random forest algorithm on the prediction of electric vehicle
  charging load,'' \emph{Energies}, vol.~11, no.~11, p. 3207, 2018.

\bibitem{lu2017load}
K.~Lu, W.~Sun, C.~Ma, S.~Yang, Z.~Zhu, P.~Zhao, X.~Zhao, and N.~Xu, ``Load
  forecast method of electric vehicle charging station using {SVR} based on
  {GA-PSO},'' in \emph{IOP Conference Series: Earth and Environmental Science},
  vol.~69, no.~1.\hskip 1em plus 0.5em minus 0.4em\relax IOP Publishing, 2017,
  p. 012196.

\bibitem{zhu2019electric}
J.~Zhu, Z.~Yang, M.~Mourshed, Y.~Guo, Y.~Zhou, Y.~Chang, Y.~Wei, and S.~Feng,
  ``Electric vehicle charging load forecasting: A comparative study of deep
  learning approaches,'' \emph{Energies}, vol.~12, no.~14, p. 2692, 2019.

\bibitem{zhang2022short}
J.~Zhang, C.~Liu, and L.~Ge, ``Short-term load forecasting model of electric
  vehicle charging load based on {MCCNN-TCN},'' \emph{Energies}, vol.~15,
  no.~7, p. 2633, 2022.

\bibitem{p2023comparative}
M.~P~Sasidharan, S.~Kinattingal, and S.~Simon, ``Comparative analysis of deep
  learning models for electric vehicle charging load forecasting,''
  \emph{Journal of the Institution of Engineers (India): Series B}, vol. 104,
  no.~1, pp. 105--113, 2023.

\bibitem{ma2022multistep}
T.-Y. Ma and S.~Faye, ``Multistep electric vehicle charging station occupancy
  prediction using hybrid lstm neural networks,'' \emph{Energy}, vol. 244, p.
  123217, 2022.

\bibitem{mohammad2023energy}
F.~Mohammad, D.-K. Kang, M.~A. Ahmed, and Y.-C. Kim, ``Energy demand load
  forecasting for electric vehicle charging stations network based on convlstm
  and biconvlstm architectures,'' \emph{IEEE Access}, vol.~11, pp.
  67\,350--67\,369, 2023.

\bibitem{wang2023short}
S.~Wang, C.~Zhuge, C.~Shao, P.~Wang, X.~Yang, and S.~Wang, ``Short-term
  electric vehicle charging demand prediction: A deep learning approach,''
  \emph{Applied Energy}, vol. 340, p. 121032, 2023.

\bibitem{unterluggauer2021short}
T.~Unterluggauer, K.~Rauma, P.~J{\"a}rventausta, and C.~Rehtanz, ``Short-term
  load forecasting at electric vehicle charging sites using a multivariate
  multi-step long short-term memory: A case study from {Finland},'' \emph{IET
  Electrical Systems in Transportation}, vol.~11, no.~4, pp. 405--419, 2021.

\bibitem{cao2024feature}
T.~Cao, Y.~Xu, G.~Liu, S.~Tao, W.~Tang, and H.~Sun, ``Feature-enhanced deep
  learning method for electric vehicle charging demand probabilistic
  forecasting of charging station,'' \emph{Applied Energy}, vol. 371, p.
  123751, 2024.

\bibitem{yu2019review}
Y.~Yu, X.~Si, C.~Hu, and J.~Zhang, ``A review of recurrent neural networks:
  {LSTM} cells and network architectures,'' \emph{Neural Computation}, vol.~31,
  no.~7, pp. 1235--1270, 2019.

\bibitem{vaswani2017attention}
A.~Vaswani, ``Attention is all you need,'' \emph{Advances in Neural Information
  Processing Systems}, 2017.

\bibitem{salih2024perspective}
A.~M. Salih, Z.~Raisi-Estabragh, I.~B. Galazzo, P.~Radeva, S.~E. Petersen,
  K.~Lekadir, and G.~Menegaz, ``A perspective on explainable artificial
  intelligence methods: Shap and lime,'' \emph{Advanced Intelligent Systems},
  p. 2400304, 2024.

\bibitem{holzinger2022explainable}
A.~Holzinger, A.~Saranti, C.~Molnar, P.~Biecek, and W.~Samek, ``Explainable
  {AI} methods-a brief overview,'' in \emph{International workshop on extending
  explainable AI beyond deep models and classifiers}.\hskip 1em plus 0.5em
  minus 0.4em\relax Springer, 2022, pp. 13--38.

\end{thebibliography}

\end{document}